\documentclass[11pt,twoside
]{article}

\usepackage{baaa2011}
\usepackage{graphicx}
\usepackage{subfigure}
\usepackage{psfrag}
\usepackage{amssymb}
\usepackage[spanish,activeacute,english]{babel}
\usepackage[latin1]{inputenc}
\usepackage[T1]{fontenc} 
\usepackage{ae,aecompl} 
\usepackage{latexsym}
\usepackage{verbatim}
\usepackage{amsmath}
\usepackage{stmaryrd}
\usepackage{amsfonts}
\usepackage{amssymb}
\usepackage{wasysym}
\usepackage[colorlinks=true,dvips]{hyperref}

\begin{document}
\myselectenglish
\vskip 1.0cm
\markboth{F. Mauro, C. Moni Bidin, D. Geisler}
{Photometric study of Galactic star clusters in the VVV survey}

\pagestyle{myheadings}
\vspace*{0.5cm}
\noindent PRESENTACI\'ON ORAL
\vskip 0.3cm
\title{Photometric study of Galactic star clusters in the VVV survey}


\author{F. Mauro,
C. Moni Bidin,
D. Geisler}

\affil{
Departamento de Astronom\'ia, Universidad de Concepci\'on, Casilla 160-C, Concepci\'on, Chile
}

\begin{abstract}
We show the preliminary analysis of some Galactic stellar clusters (GSCls) candidates and the results of the analysis of two new interesting GSCls found in the ``VISTA Variables in the V\'ia L\'actea'' (VVV) Survey.
The VVV photometric data are being used also to improve the knowledge of the Galactic structure.
The photometric data are obtained with the new automatic photometric pipeline VVV-SkZ\_pipeline.

\end{abstract}
\begin{resumen}
Mostraremos el an\'alisis preliminar de algunos candidatos a c\'umulos Gal\'acticos estelares (GSCl) y los resultados de los an\'alisis de dos  nuevos interesantes GSCls encontrados en ``VISTA Variables in the V\'ia L\'actea'' (VVV) Survey.
Los datos fotom\'etricos VVV estan siendo usados tambien para mejorar el conocimiento de la estructura Gal\'actica.
Los datos fotom\'etricos fueron obtenidos con la nueva ``pipeline'' fotom\'etrica autom\'atica VVV-SKZ\_pipeline.
\end{resumen}

\section{Introduction}

Galactic stellar clusters (GSCls) are the perfect laboratories for studying a wide variety of fundamental problems in stellar and galactic astrophysics.
They are testbeds for the understanding of stellar dynamics and evolution.
They are tracers of the structure, formation and chemical evolution of the Galaxy and its distinct components.
It is well studied that the majority of stars with mass $M>0.5\;M_\odot$ form in clustered environment (Lada \& Lada 2003).

The discovery of new Galactic globular clusters (GGlCl) is important, since they serve as dynamical probes of the Galaxy's complex kinematics and interaction history, and are cornerstones of the distance scale.

The ``Vista Variables in the V\'ia L\'actea'' (VVV) Public Survey (Minniti et al. 2010; Saito et al. 2010; Catelan et al. 2011; Saito et al. 2012) is gathering near-IR ($YZJHK_s$) data of the Galactic Bulge ($-10\leq l\leq+10$, $-10\leq b\leq +5$) and the adjacent part of inner disk ($-65\leq l\leq-10$, $-2\leq b\leq +2$).
It includes 36 known GGlCls and more than 300 open clusters (OpCls).
Up to now two new GGlCls and a hundred of OpCls were discovered using VVV data (Minniti et al. 2011; Borissova et al. 2011; Moni Bidin et al. 2011).

The analysis of Galactic stellar clusters (GSCls) candidates is not the only important work based on VVV photometry to improve our knowledge of the galactic structure.
VVV PSF-fitting photometry of known GSCls is used to improve the distance of various GSCls (Majaess et al. 2012, in prep) to determine the distance of Cepheids associated to them or their association with the cluster (Majaess et al. 2011, TW Nor and Lyng\aa{} 6).
The same technique can also be applied to planetary nebulae and this improved knowledge of the distance of GSCls  be used to trace better the Galactic arms.

The photometric data are obtained with the new automatic photometric pipeline VVV-SkZ\_pipe\-line (Mauro et al. 2012, PASP, submitted).
The final catalogs were calibrated using the 2MASS catalog (Skrutskie et al. 2006) as standard catalog.

\section{Photometric Analysis of the candidates}

We use a preliminary nomenclature for the candidates, following the one introduced by Minniti et al. (2011), Borissova et al. (2011), and Moni Bidin et al. (2011).

We used a decontamination procedure based on the method of Gallart et al. (2003).
For each star in the comparison field, it finds the nearest star located at the $K_s-(J-K_s)$ plane within a given maximum distance, and rejects it.
Stars in the comparison field without a counter-part in the object area are flagged as ``unmatched''.
More than one comparison field was used to assure a good result of the decontamination.


In this work we show the preliminary results of the photometric analysis of the new candidates.
For the analysis of the new GSCls VVV-CL002 and VVV-CL003, here we only note their most interesting characteristic: the first one appears to be one of the innermost GGlCls, while the second one seems be the first GSCl located beyond the Galactic Bulge.
We refer to Moni Bidin et al. (2011) for more detailed information.

~

\begin{figure}[!ht]
\centering
\includegraphics[width=.45\textwidth]{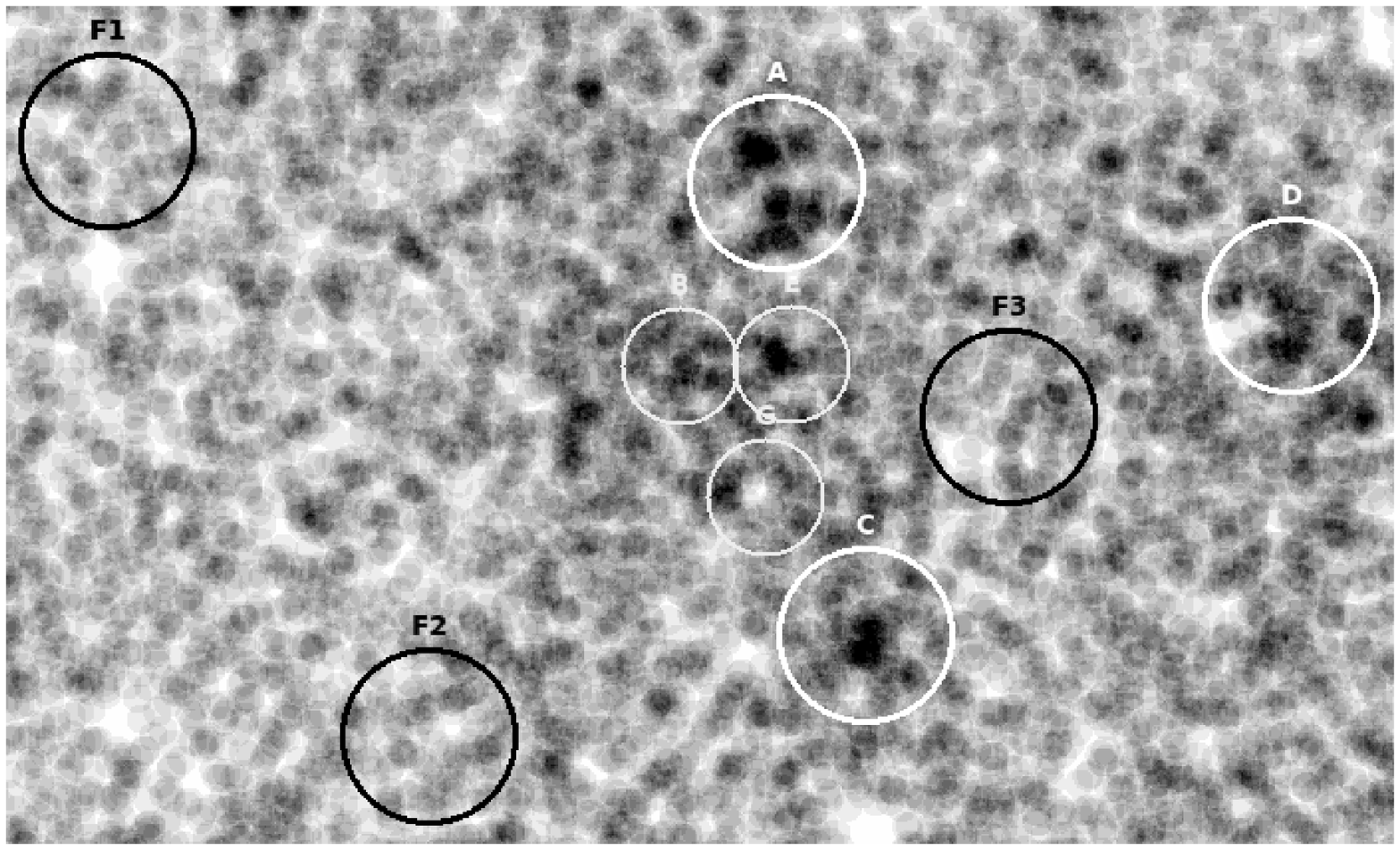}
\includegraphics[width=.35\textwidth]{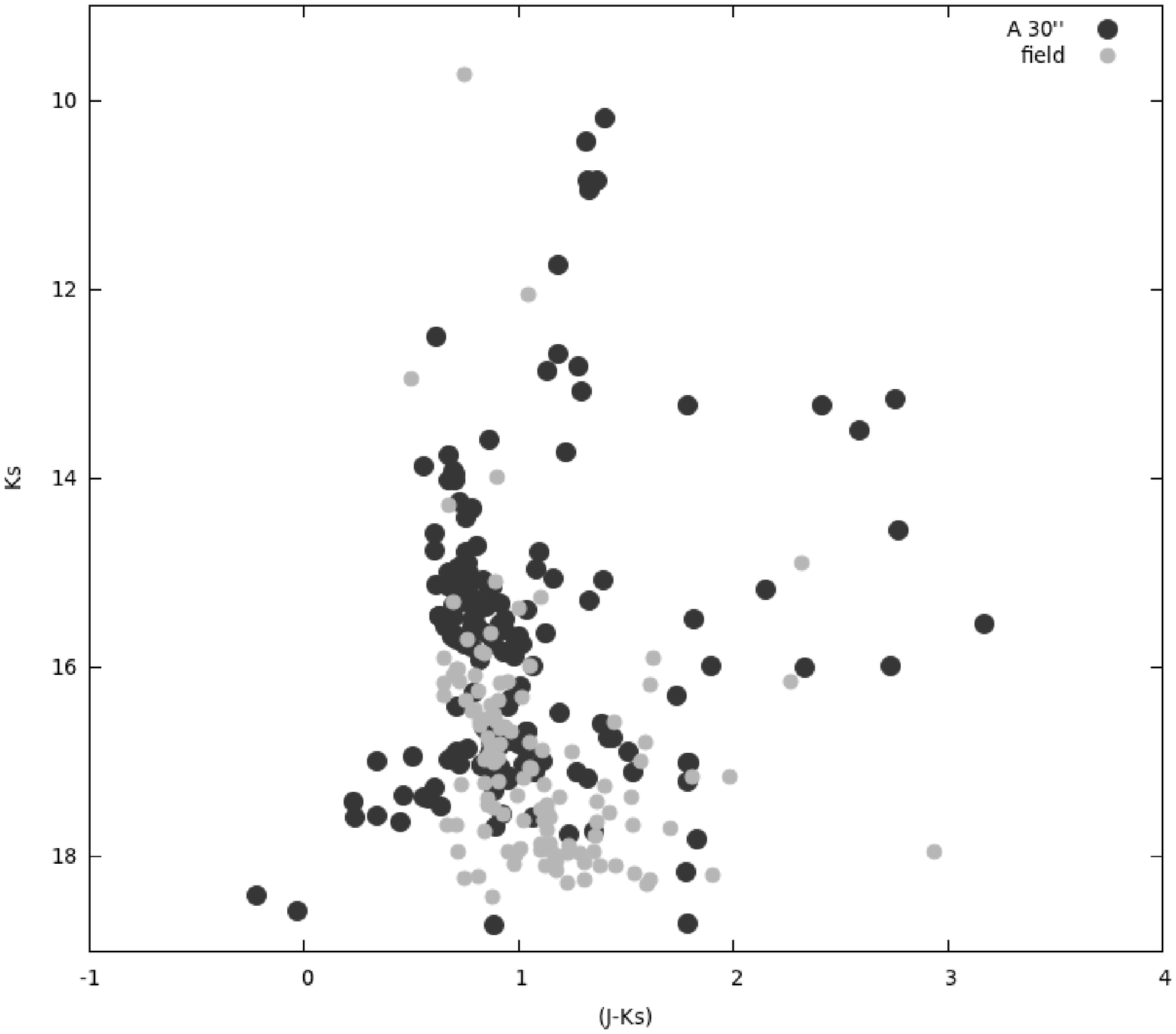}
\includegraphics[width=.35\textwidth]{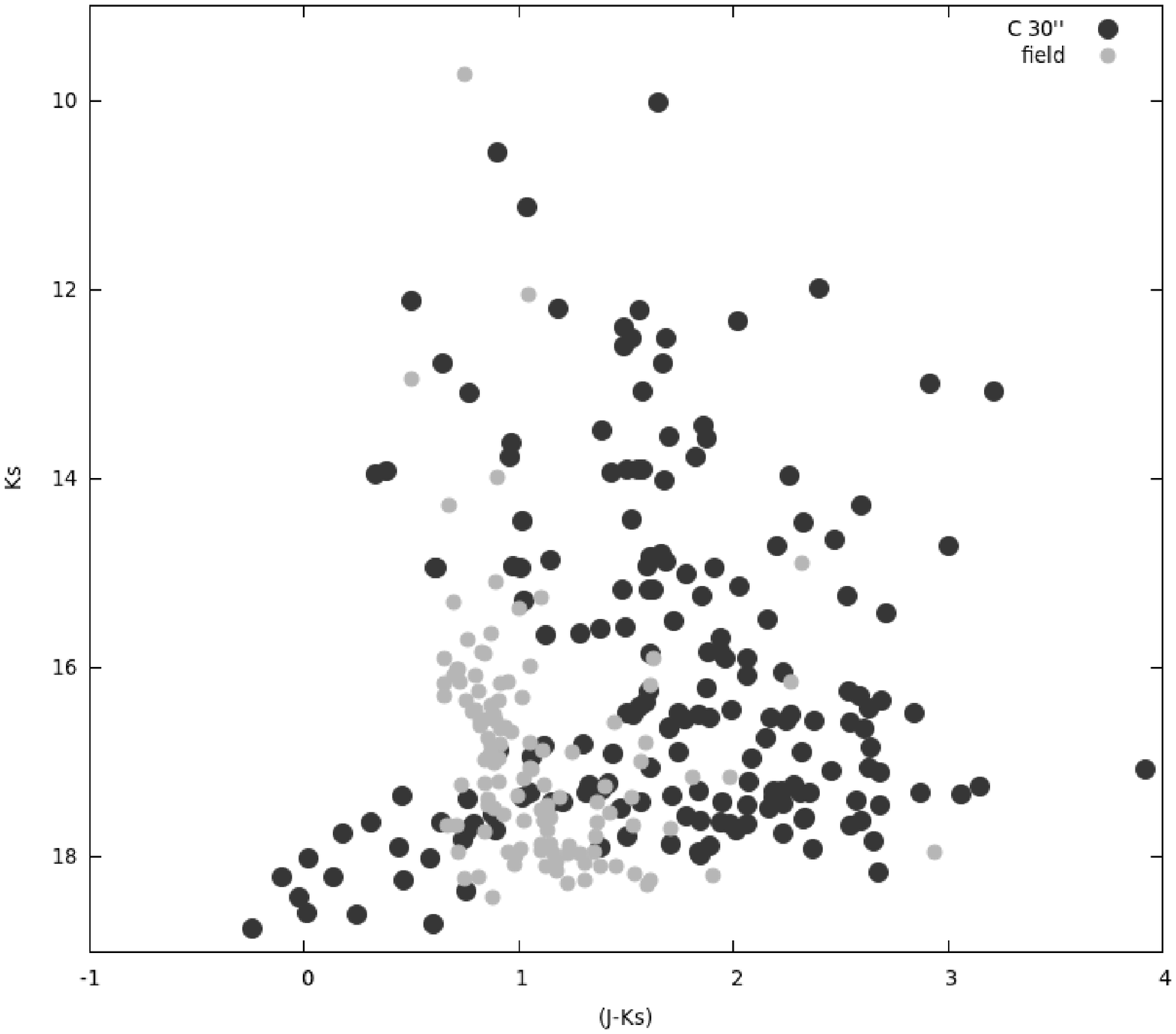}
\includegraphics[width=.35\textwidth]{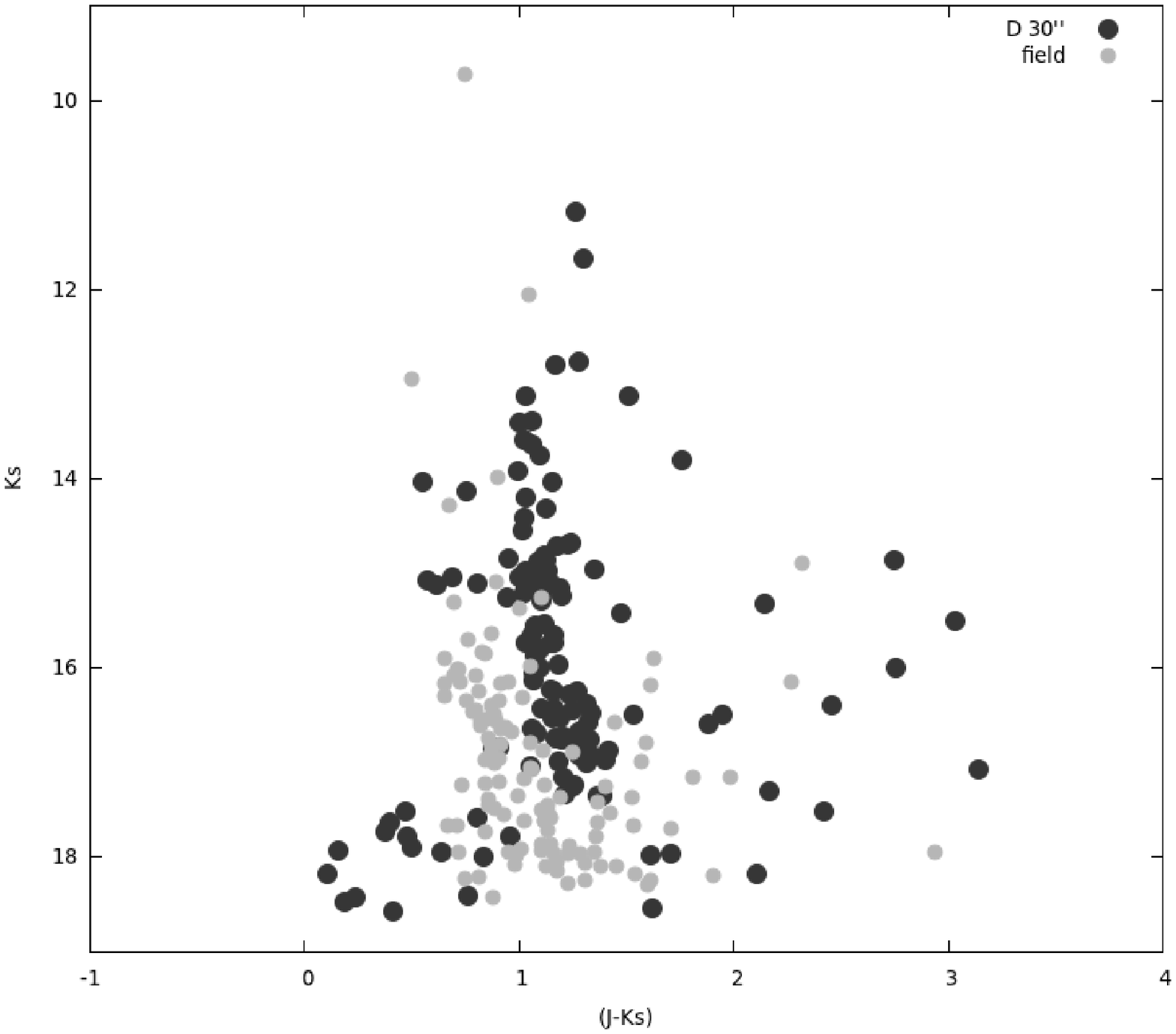}
\caption{{\it Top left:} Density map of the stars with magnitude $K_s\le16.0$ for the area of CL101. The three main over-densities A (top), C (bottom) and D (right) and the comparison fields are marked with circles of 30'' radius. {\it Top right and bottom:} Decontaminated color-magnitude diagrams for the three main over-densities with over-plotted a comparison field.}\label{Mauro-Francesco-fig1-2-3-4}
\end{figure}

{\bf CL101 ~}
The candidate CL101 is located in $l\simeq298.550\; b\simeq-0.162$ (tile d041, Saito et al. 2012).
It appears to be a complex system of several Stellar Associations.
In Figure \ref{Mauro-Francesco-fig1-2-3-4} in the density map (a product of the VVV-SkZ\_pipeline) the three main overdensity are marked with a 30''-radius circles.
The overdensity marked as A is the already discovered, but poorly studied, stellar association (StAs) FSR 1606/Alessi 52 (Dias et al. 2002), while the other 2 main overdensities C and D were unknown.
A few acrminutes to the east is also located  Ruprecht 102 (Kharchenko et al. 2005), a known, but poorly studied, open cluster.
We will use the parameters of the stellar clusters to check if they are coeval and form a unique stellar association.

~

\begin{figure}[!ht]
\centering
\includegraphics[width=.45\textwidth]{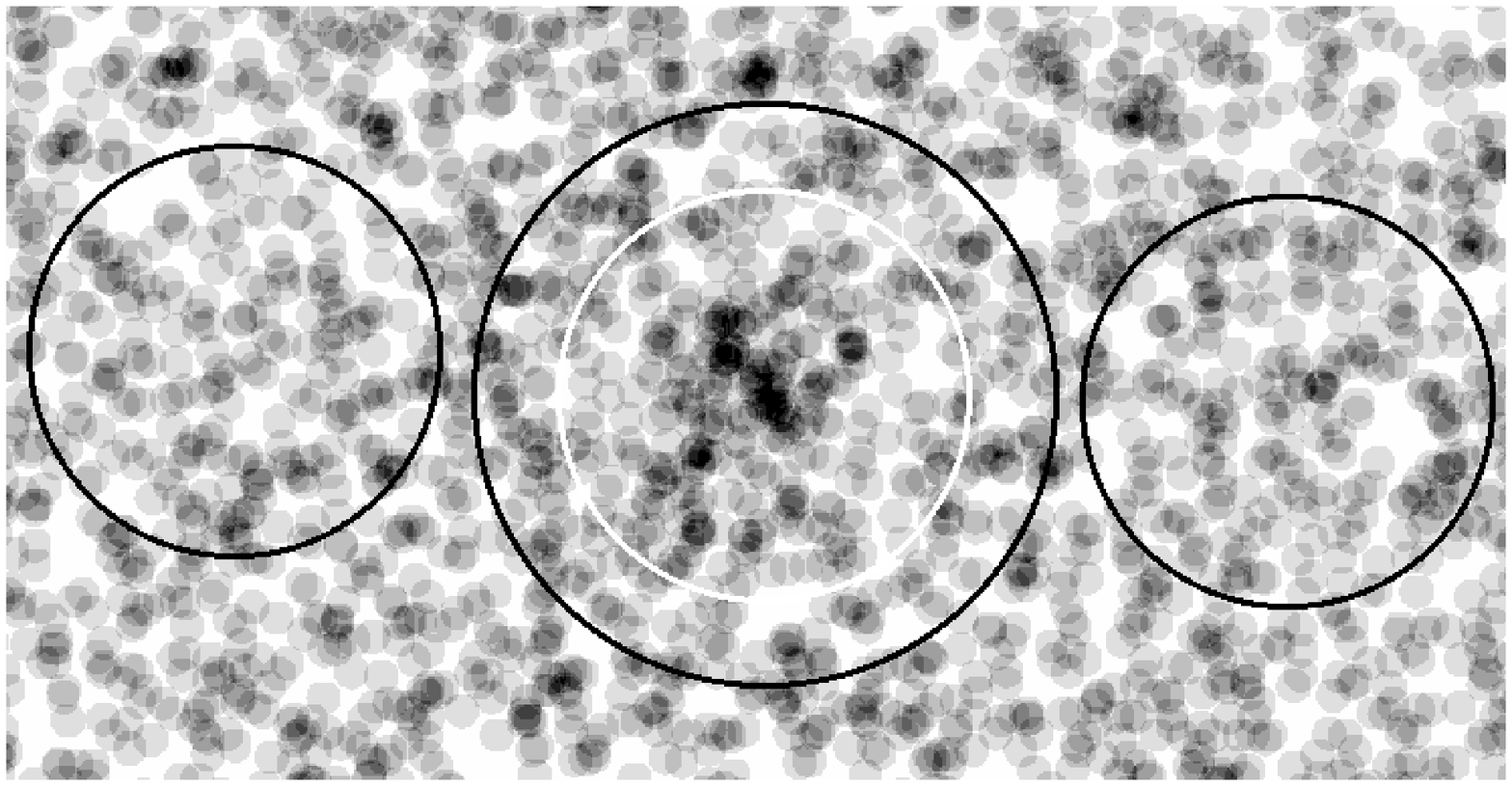}~
\includegraphics[width=.32\textwidth]{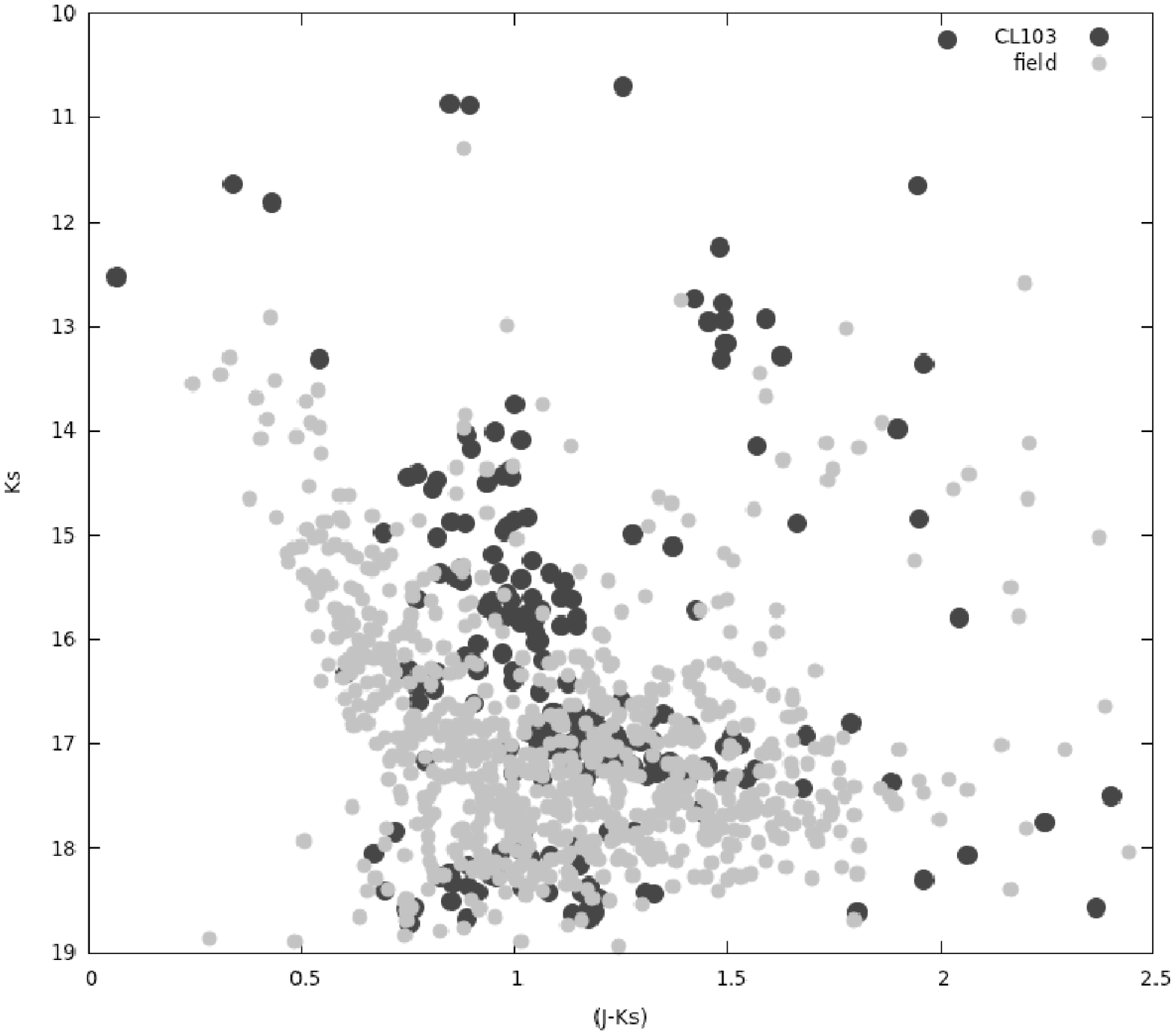}
\caption{Density map of the stars with magnitude $K_s\leq16.5$ for the area (white circle: CL103, black circles: comparison fields;  radius 1') and decontaminated color-magnitude diagram with over-plotted a comparison field.} \label{Mauro-Francesco-fig5-6}
\end{figure}

{\bf CL103 ~}
The candidate CL103 is located in $l\simeq295.476\; b\simeq0.074$ (tile d077).
It is a compact object (see Figure \ref{Mauro-Francesco-fig5-6}) with core radius and half-light radius of about $20 \arcsec$ and tidal radius of $60 \arcsec$.
Its decontaminated color-magnitude diagram (CMD) is not well defined, but the top part of its main sequence (MS) and red-giant clump (RGC) are definitely well separated in the comparison with all the four used comparison fields, localized within one-two arcminutes away.

~

\begin{figure}[!ht]
\centering
\includegraphics[width=.35\textwidth]{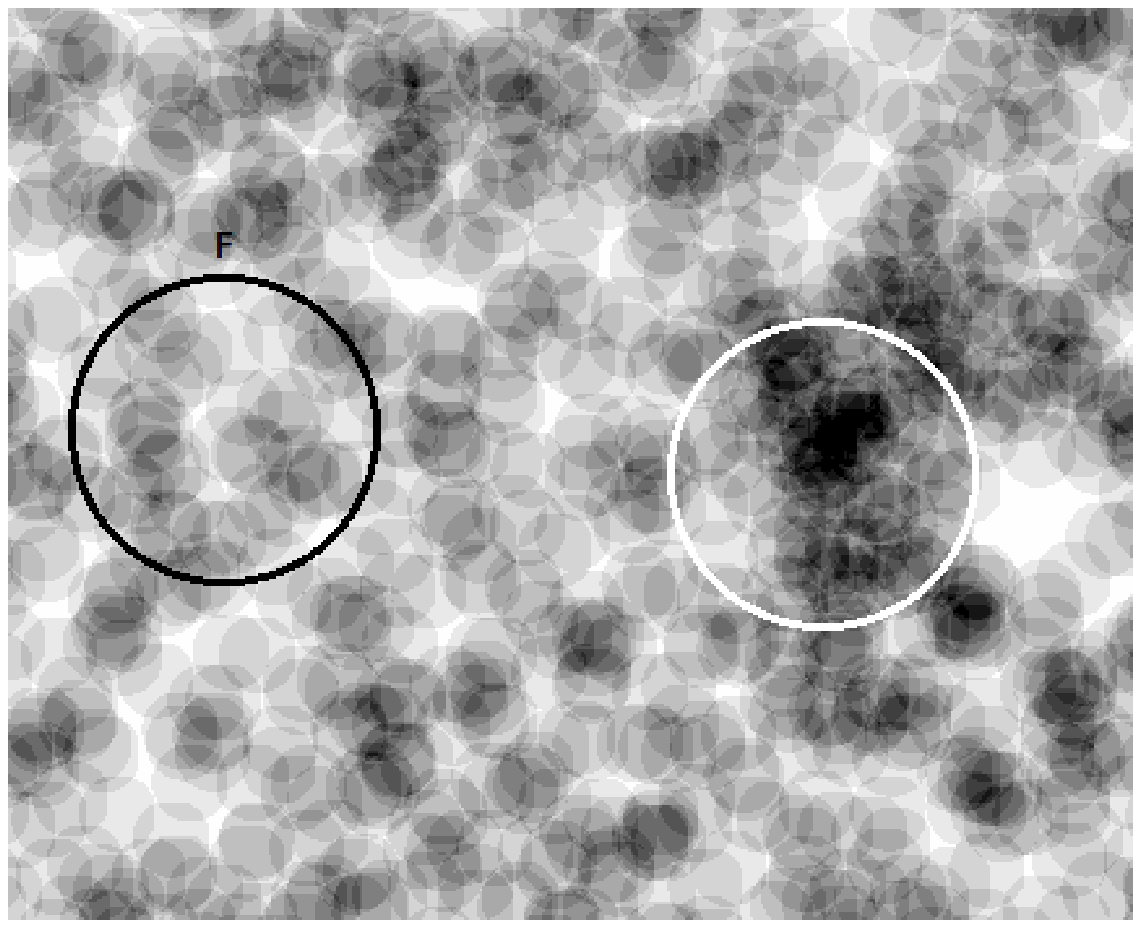}~
\includegraphics[width=.33\textwidth]{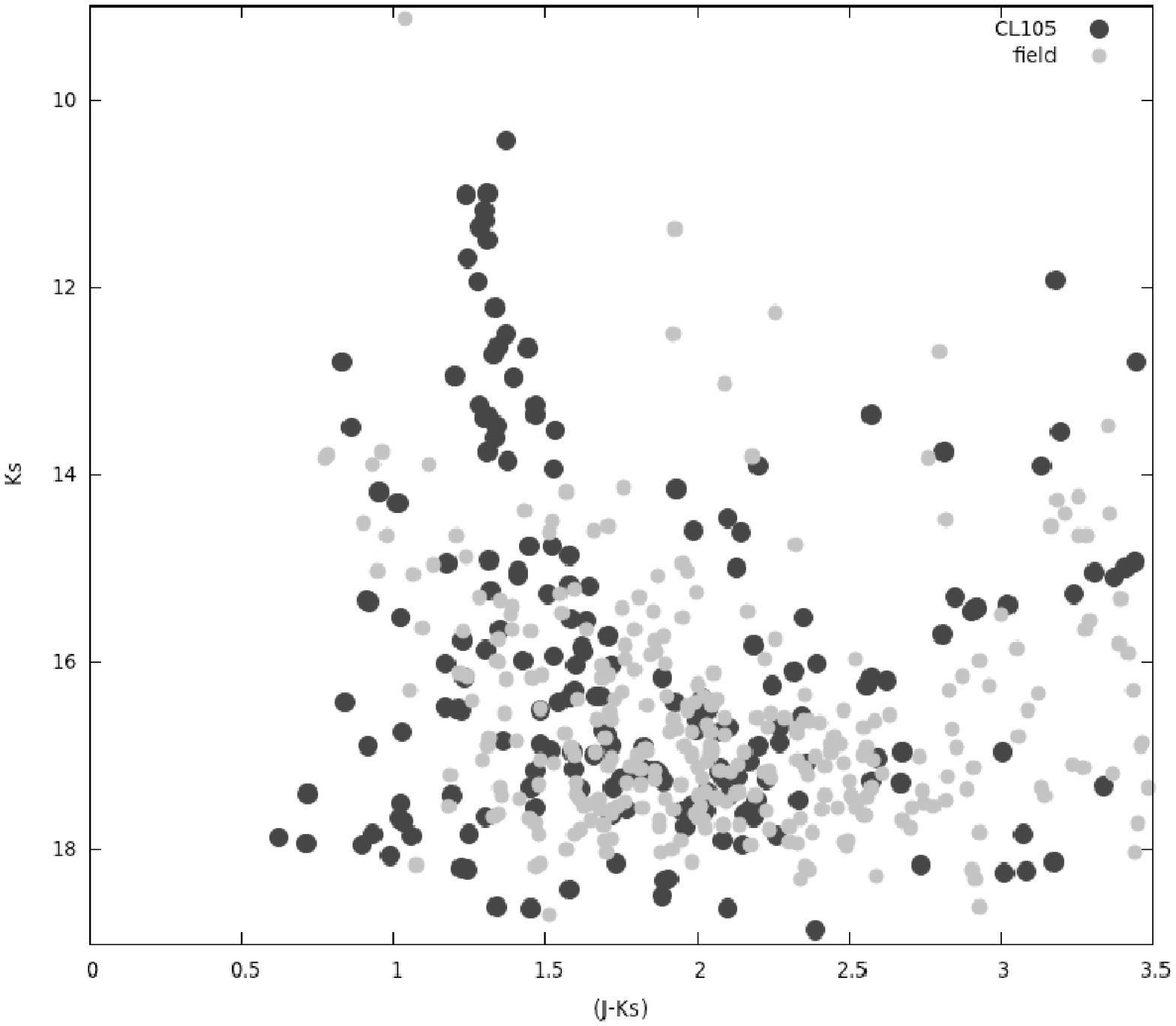}
\caption{Density map of the stars with magnitude $K_s\leq16.5$ for the area (white circle: CL105, black circle: comparison field;  radius 0.5') and decontaminated color-magnitude diagram with over-plotted a comparison field.} \label{Mauro-Francesco-fig7-8}
\end{figure}

{\bf CL105 ~}
The candidate CL105 is located in $l\simeq330.035\;b=\simeq0.751$ (tile d101).
It is a small object (see Figure \ref{Mauro-Francesco-fig7-8}) with core radius and half-light radius of about $12 \arcsec$ and tidal radius of $30 \arcsec$.
The decontaminated CMD shows a sequence characteristic of a young object.
The brighter part of the MS of CL105 is located in an area of the CMD poorly populated in the CMD of the several chosen comparison fields.

\section{Conclusions}
In this paper we presented the preliminary analysis of three new stellar-cluster candidates, found in the ``VVV - Vista Variables in the V\'ia L\'actea'' Survey.
All the three candidates present a decontaminated color-magnitude diagram (CMD) showing characteristics that suggest they could be stellar clusters.
CL101 is the most interesting candidate since, presenting several likely stellar clusters in several tenth of square arcminutes, it could have been a active star-forming area. We aim to  study it further to determine if it is a single stellar association.

\acknowledgements
We gratefully acknowledge use of data from the ESO Public Survey program ID~179.B-2002 taken with the VISTA telescope, and data products from the Cambridge Astronomical Survey Unit. We acknowledge support by the FONDAP Center for Astrophysics 15010003, BASAL Center for Astrophysics and Associated Technologies PFB-06/2007. This investigation made use of data from the Two Micron All Sky Survey.

\begin{referencias}
\reference Borissova, J., Bonatto, C., Kurtev, R., et al. 2011, A\&A, 532A, 131
\reference Catelan, M., Minniti, D., Lucas, P. W., et al. 2011, RR Lyrae Stars, Metal-Poor Stars, and the Galaxy, 145
\reference Dias W. S., Alessi B. S., Moitinho A. \& L\'epine J. R. D., 2002, A\&A 389, 871
\reference Emerson, J., \& Sutherland, W. 2010, The Messenger, 139, 2
\reference Gallart, C., Zoccali, M., Bertelli, G., et al. 2003, AJ, 125, 742
\reference Irwin, M.J., Lewis, J., Hodgkin, S., et al. 2004, SPIE, 5493, 411
\reference Kharchenko, N., V.,  Piskunov, A. E., R\"oser, S., et al. 2005, A\&A 438, 1163
\reference Lada, C.J., \& Lada, E.A. 2003, ARA\&A, 41, 57
\reference Majaess, D., Turner, D., Moni Bidin, C., et al. 2011, ApJ, 741L, 27 
\reference Minniti, D., Lucas, P.W., Emerson, J.P., et al. 2010, New A., 15, 433
\reference Minniti, D., Hempel, M., Toledo, I., et al. 2011, A\&A, 527, 81
\reference Moni Bidin, C, Mauro, F., Geisler, D., et al. 2011, A\&A, 535A, 33 
\reference Saito, R., Hempel, M., Alonso-Garc\'{\i}a, J., et al. 2010, The Messenger, 141, 24
\reference Saito, R., Hempel, M., Minniti, D., et al. 2012, A\&A, 537A, 107 
\reference Skrutskie,M.F., et al. 2006, AJ, 131, 1163 
\end{referencias}

\end{document}